\documentclass[preprint,aps,pre,showpacs,superscriptaddress]{revtex4}
\usepackage{graphicx}
\usepackage{bm}
\usepackage{amssymb}

\begin{document}
\title{Elasticity of polymer vesicles by osmotic pressure:
an intermediate theory between fluid membranes and solid shells}
\author{Z. C. Tu}
\affiliation{Institute of Theoretical Physics, Academia Sinica,
P.O.Box 2735 Beijing 100080, China}
\author{L. Q. Ge}\affiliation{State Key Laboratory of Bioelectronics, Southeast University, Nanjing 210096, China}
\affiliation{International Joint Lab, Key Lab of Colloid and
Interface Science, The Center for Molecular Science, Institute of
Chemistry, Academia Sinica, Beijing 100080, P.R.China}
\author{J. B. Li}\affiliation{International Joint Lab, Key Lab of Colloid and
Interface Science, The Center for Molecular Science, Institute of
Chemistry, Academia Sinica, Beijing 100080, P.R.China}
\author{Z. C. Ou-Yang}\email[Email address: ] {oy@itp.ac.cn}
\affiliation{Institute of Theoretical Physics, Academia Sinica,
P.O.Box 2735 Beijing 100080, China} \affiliation{Center for
Advanced Study, Tsinghua University, Beijing 100084, P.R.China}
\begin{abstract}
The entropy of a polymer confined in a curved surface and the
elastic free energy of a membrane consisting of polymers are
obtained by scaling analysis. It is found that the elastic free
energy of the membrane has the form of the in-plane strain energy
plus Helfrich's curvature energy [Z. Naturforsch. C \textbf{28},
693 (1973)]. The elastic constants in the free energy are obtained
by discussing two simplified models: one is the polymer membrane
without in-plane strains and asymmetry between its two sides,
which is the counterpart of quantum mechanics in curved surface
[Jensen and Koppe, Ann. Phys. \textbf{63}, 586 (1971)]; another is
the planar rubber membrane with homogeneous in-plane strains. The
equations to describe equilibrium shape and in-plane strains of
the polymer vesicles by osmotic pressure are derived by taking the
first order variation of the total free energy containing the
elastic free energy, the surface tension energy and the term
induced by osmotic pressure. The critical pressure, above which
spherical polymer vesicle will lose its stability, is obtained by
taking the second order variation of the total free energy. It is
found that the in-plane mode also plays important role in the
critical pressure because it couples with the out-of-plane mode.
Theoretical results reveal that polymer vesicles possess the
mechanical properties intermediate between fluid membranes and
solid shells.
\end{abstract}
\pacs{61.41.+e} \maketitle
\section{\label{introd} introduction}
Thin thickness structures exist widely in Nature. Many things such
as eggs, snails, airplanes and so on in our daily life are covered
with solid shells that play protective roles. In the realm that we
cannot see with naked eyes, virus usually have protein shells, and
eukaryotic cells are enclosed by cell membranes that consist of
lipids, proteins and carbohydrates etc. A lipid molecule has a
polar hydrophilic head group and one or two hydrophobic
hydrocarbon tails. When a quantity of lipid molecules disperse in
water, they will assemble themselves into a lipid bilayer in which
the hydrophilic heads shield the hydrophobic tails from the water
surroundings because of the hydrophobic forces. Solid shells and
lipid bilayers are, respectively, in the categories of hard
condensed matter and soft one. Their mechanical properties have
attracted much attention for a long time
\cite{landau,Pogorelov,Helfrich,Reinhard,SeifertU,oytsf}.

The significant difference between solid shells and lipid bilayers
is that the former can endure the in-plane shear stress but the
latter cannot. Due to this difference, solid shells and lipid
bilayers have different forms of deformation energy. Under the
assumption of homogenous and isotropic bulk materials and in the
limit of thin thickness, the elastic free energy per unit area of
a solid shell is expressed as \cite{landau}:
\begin{eqnarray}
\label{solidsh} \mathcal{E}_{sh}&=&\frac{D}{2} \left[
(2H)^2-2(1-\nu)K \right]\nonumber\\
&+&\frac{C}{2(1-\nu^2)}\left[(2J)^2-2(1-\nu)Q \right]
,\end{eqnarray} where $D=(1/12)Yh^3/(1-\nu ^2)$ and $C=Yh$ are
bending rigidity and in-plane stiffness of the shell. $Y$, $\nu$
and $h$ are, respectively, the Young's modulus, the Poisson ratio
and the thickness of the shell. $2J$ and $Q$ are the trace and
determinant of the in-plane strain tensor, respectively. For a
spherical solid shell with radius $R$, the critical osmotic
pressure (i.e., the pressure difference between the outer surface
and inner one of the shell, above which the shell loses its
stability) is \cite{Pogorelov}:
\begin{equation}
p_{cs}=\frac{2Yh^2}{\sqrt{3(1-\nu^2)}R^2}.\label{critpshell}\end{equation}

In 1973, Helfrich \cite{Helfrich} recognized that the lipid
bilayer was just like liquid crystal in smectic A phase at room
temperature. Based on the elastic theory of liquid crystal
\cite{degennes}, he proposed the curvature energy per unit area of
the bilayer
\begin{equation}\label{Helfrich} \mathcal{E}_{lb}=(k_c/2)(2H+c_0)^2+\bar{k}K, \end{equation}
where $k_c$ and $\bar{k}$ are elastic constants; and $H$, $K$ and
$c_0$ are mean curvature, Guassian curvature and spontaneous
curvature of the lipid bilayer, respectively. For phospholipid
bilayers at room temperature $T$, the persistence length is
usually much larger than the size of the membranes and the effect
of shape fluctuations is negligible \cite{Reinhard} because of
$k_c\approx 10^{-19}$J$\gg k_BT$ \cite{Mutz2}, where $k_B$ is the
Boltzmann factor. The free energy of a closed bilayer under the
osmotic pressure $p$ is written as $\mathcal{F}_{lb}=\oint
(\mathcal{E}_{lb}+\lambda) dA+p\int dV$, where $dA$ is the area
element and $V$ the volume enclosed by the closed bilayer.
$\lambda$ is the surface tension of the bilayer. The first order
variation of $\mathcal{F}_{lb}$ gives the shape equation of closed
bilayer \cite{oypr}:{\small
\begin{equation}\label{shapelb}p-2\lambda
H+k_c\nabla^2(2H)+k_c(2H+c_0)(2H^2-c_0H-2K)=0.\end{equation}}For a
spherical lipid bilayer with radius $R$, the critical osmotic
pressure for stability is \cite{oybook}:
\begin{equation}
p_{cl}=\frac{2k_{c}(6-c_{0}R)}{R^{3}}.\label{critplipd}\end{equation}
It follows that $p_{cl}\sim k_{c}/R^{3}$ because the typical value
of $c_{0}R$ is about 1. Therefore, lipid bilayer is, indeed, much
softer than solid shell.

Are there membranes intermediate in the state between Helfrich's
fluid lipid bilayers and classical solid shells? The polymer
vesicles discussed below may be as a example. In the last decade,
Decher invented the layer-by-layer assembling technique
\cite{Decher}. Following this technique, Caruso \emph{et al.} made
spherical polyelectrolyte capsules by the step-wise adsorption of
polyelectrolytes onto charged colloidal templates and then
decomposition of the templates \cite{Caruso,Donath}. The capsules
composed of about 10 layers of alternating polystyrene sulfonate
and polyallylamine hydrochloride. The thickness $h$ of capsules
was about tens of nanometers which was remarkably less than their
radii $R$ (several micrometers). Gao \emph{et al.} \cite{cgao}
found that the spherical polyelectrolyte capsule lost its
stability and changed its shape abruptly above some threshold of
osmotic pressure $p_c$ which was proportional to $R^{-2}$ and
$h^2$. In their experiment, the thickness dependence of $p_c$
might not be exact because the polyelectrolyte capsule with more
than 10 layers was chemically instable as they claimed. They also
explained their results through the stability theory of classical
elastic solid shells \cite{landau,Pogorelov}. But it is well known
that the classical theory is based on the assumption of homogenous
and isotropic bulk materials which entirely ignores the
characteristic of the polyelectrolyte capsule consisting of many
polymers. If considering the polymer structures of spherical
polyelectrolyte capsule, can we still derive $p_c\sim R^{-2}$?

In this paper, we will answer above questions. To do that, we
derive the entropy of a polymer confined in a curved surface and
the elastic free energy of a membrane consisting of polymers by
scaling analysis. It is found that the elastic free energy of the
polymer membrane has the form of the in-plane strain energy plus
Helfrich's curvature energy. The elastic constants in the free
energy are obtained by discussing two simplified models: one is
the polymer membrane without in-plane strains and asymmetry
between its two sides, which is the counterpart of quantum
mechanics in curved surface \cite{Jensen}; another is the planar
rubber membrane with homogeneous in-plane strains. The equations
to describe equilibrium shape and in-plane strains of polymer
vesicles by osmotic pressure are derived by taking the first order
variation of the total free energy containing the elastic free
energy, the surface tension energy and the term induced by osmotic
pressure. The critical pressure, above which a spherical polymer
vesicle will lose its stability, is obtained by taking the second
order variation of the total free energy. It is found that the
in-plane mode also plays important role in the critical pressure
because it couples with the out-of-plane mode. These theoretical
results reveal that polymer vesicles possess the mechanical
properties being intermediate between Helfrich's fluid membranes
and classical solid shells.

The following contents of this paper are organized as below: In
Sec. \ref{frgsec}, we derive the free energy of polymer membrane
by using the scaling concepts \cite{gennes}. In Sec.
\ref{shsteqsec}, we obtain the shape and in-plane strain equations
of closed polymer vesicles by using the surface variation theory
developed in Ref.\cite{tzcpre,tzcjpa}. In Sec. \ref{stabsec}, we
discuss the mechanical stability of the spherical polymer vesicle
by taking the second order variation of the free energy. In
Sec.\ref{Conclusion}, we give a brief summary and discussion.

\section{\label{frgsec} The free energy of polymer membrane}
The polymer membrane concerned in this paper is one or a few thin
layers consisting of cross-linking polymer structure like rubber
\cite{Treloar} at molecular levels. It can be represented as a
mathematical surface with curvature and strains. It is hard to
derive its free energy in a strict way. But we can drive it by
using scaling concepts in polymer physics proposed by de Gennes
\cite{gennes}. In the following contents, we take de Gennes'
convention: the entropy $S$ is a dimensionless quantity and
Boltzmann factor $k_B$ is implicated in temperature $T$.

\subsection{The free energy of polymer membrane}
If we take the Gaussian chain model \cite{Dio}, the root of mean
square end-to-end distance of a polymer is $R_0\sim \sqrt{N}b_0$,
where $b_0$ is segment length of the polymer and $N$ is the number
of segments. Assume that the principal radii of the surface are
much larger than $R_0$. If the in-plane strain tensor of the
surface is denoted by ${\bm\epsilon}$ which is assumed to be a
small quantity, the entropy of the polymer confined the surface
must be the function of $2HR_0$, $KR_0^2$, $2J$ and $Q$ because it
is a dimensionless invariant quantity under the transformation of
coordinates, where $H$, $K$, $J=\mathrm{tr}{\bm\epsilon}$ and
$Q=\mathrm{tr}{\bm\epsilon}$ are the mean curvature of the
surface, the Gaussian curvature of the surface, the trace of
strain tensor and the determinant of strain tensor, respectively.
Thus we can expand it as
\begin{equation}S\sim A_1(2HR_0)+A_2(2HR_0)^2+A_3KR_0^2+B_2(2J)^2+B_3Q\end{equation}
up to the second order terms, where $A_1, A_2, A_3, B_2, B_3$ are
constants. In this expression, an unimportant constant term is
neglected. Moreover, one must notice that generally we have $K\sim
(2H)^2$ and $Q\sim (2J)^2$. Additionally, there is no first order
term of $2J$ in the expression of the entropy because we expect
that $-{\bm\epsilon}$ plays the same role as ${\bm\epsilon}$ in
the entropy. It is useful to write the entropy in another
equivalent form \begin{equation}S \sim
A_{2}R_{0}^{2}(2H+c_{0})^{2}+A_{3}R_{0}^{2}K+B_{2}(2J)^{2}+B_{3}Q,\end{equation}
where $c_0=A_1/(2A_2)$ is a constant, so called spontaneous
curvature, which is expected to satisfy $|c_0R_0|\ll 1$. In fact,
$c_0$ vanishes if there is no asymmetric factor between two sides
of the surface because $H$ turn into $-H$ if we change the normal
direction of the surface.

Assume $h$ to be the membrane thickness and $M$ the number of
polymers per volume. Additionally, we neglect the the entanglement
of polymers. Consequently, the free energy per unit area of a
membrane consisting of polymers has the following form
\begin{eqnarray}
\mathcal{E}_{pm}&=&-(Mh)TS\nonumber\\
&=&\frac{k_{d}}{2}[(2J)^{2}-\mu
Q]+\frac{k_{c}}{2}(2H+c_{0})^{2}-\bar{k}K,\label{elastic}
\end{eqnarray}
where $k_d=-2B_2MhT$, $\mu=-B_3/B_2$, $k_c=-2MhTA_{2}R_{0}^{2}$,
$\bar{k}=MhTA_{3}R_{0}^{2}$.

Obviously, Eq.(\ref{elastic}) is degenerated to Helfrich's
curvature energy of fluid membranes if $k_d=0$, and to the elastic
energy of classical solid shells if $c_0=0$, $k_d=C/(1-\nu^2)$,
$k_c=D$, $\mu=2(1-\nu)$ and $\bar{k}=D(1-\nu)$.

\subsection{The elastic constants $k_d$, $\mu$ and $k_c$}
$k_d$, $\mu$, $k_c$ and $\bar{k}$ are unknown universal constants
independent of the detailed shape and the small in-plane strains
of the polymer membrane. If only discussing the closed polymer
vesicles in this paper, we need not to know $\bar{k}$ because the
integral of $\bar{k}K$ is an unimportant constant. To determine
$k_d$, $\mu$ and $k_c$, we will discuss two ideal cases: One is
the cylinder polymer membrane without any strain and asymmetry
between its two sides; another is the planar membrane with the
homogenous in-plane strains.

In the former case, we denote $\rho$ the radius of the cylinder.
On the one hand, (\ref{elastic}) is simplified as
\begin{equation}\label{cylinder1}
\mathcal{E}_{pm}=\frac{k_{c}}{2\rho^2}.
\end{equation}
On the other hand, we know there is a 1-1 correspondence relation
between polymer statistics and quantum path integral method
\cite{Dio} as shown in Table \ref{correspond}. In 1971, Jensen and
Koppe dealt with the quantum mechanics of a particle constrained
in a curved surface and obtained a nontrivial conclusion
\cite{Jensen}: the constraint would induce an effective potential
$V_{ef}=-\frac{\hbar^2}{8m}[(2H)^2-4K]$ in Shr\"{o}dinger
equation, where $m$ is the particle mass. In terms of the
correspondence rules in Table \ref{correspond}, there will be an
effective potential $U_{ef}=\frac{b_0^2}{24\beta}[(2H)^2-4K]$ for
a polymer confined in curved surface. Especially,
$U_{ef}=\frac{Tb_0^2}{24\rho^2}$ for the cylinder with radius
$\rho$. We must pay more attention to the fact that there is a
minus symbol in the potential term when we use the correspondence
rules. In fact, Yaman \emph{et al.} overlooked this fact in recent
literature \cite{Yaman}. But this flaw can not diminish the value
of their pioneer work in the study of polymer confined in curved
surface. Thus their results can be safely transplanted only if we
change the sign. Consequently, we obtain the free energy of the
cylindrical membrane consisting of $Mh$ polymers per unit area
\begin{equation}\label{cylindf}\mathcal{E}_{pm}=\frac{MhTNb_0^2}{24\rho^2}=\frac{MhTR_0^2}{24\rho^2}\end{equation}
if neglecting the entanglement between polymers. Comparing
Eq.(\ref{cylinder1}) with Eq.(\ref{cylindf}), we obtain
$k_c=MhR_0^2T/12$.

In the latter case, $H$, $K$ and $c_0$ are vanishing for planar
membrane with symmetry between its two sides. On the one hand,
(\ref{elastic}) is simplified as
\begin{equation}\label{planar}
\mathcal{E}_{pm}=\frac{k_{d}}{2}[(2J)^{2}-\mu Q].
\end{equation}
For homogenous stain ${\bm \epsilon}$, we can express it by its
components $\epsilon_{11}$, $\epsilon_{22}$ and
$\epsilon_{12}=\epsilon_{21}=0$ in some orthonormal coordinate
system so that $2J=\epsilon_{11}+\epsilon_{22}$ and
$Q=\epsilon_{11}\epsilon_{22}$.

On the other hand, we notice that there might be cross-linking
joints between polymers in the membrane. This character suggests
that the membrane should have the elastic properties of rubber
materials. In terms of the elasticity theory of rubber
\cite{Treloar}, the deformation energy of a planar rubber per area
can be expressed as
$f_r=(MhT/2)[\lambda_1^2+\lambda_2^2+1/(\lambda_1^2\lambda_2^2)-3]$,
where $\lambda_1=1+\epsilon_{11}$ and $\lambda_2=1+\epsilon_{22}$
are extensions. For small strains, it is expanded to the lowest
order terms as \begin{equation}\mathcal{E}_{pm}=
2MhT(\epsilon_{11}^2+\epsilon_{11}\epsilon_{22}+\epsilon_{22}^2)
=2MhT[(2J)-Q].\label{planar2}\end{equation} Thus we can obtain
$k_d=4MhT$ and $\mu=1$ by comparing Eq.(\ref{planar}) with
Eq.(\ref{planar2}).

Therefore, we obtain the free energy of a closed polymer vesicle
under osmotic pressure $p$:
\begin{equation}\label{freeeng}
\mathcal{F}=\oint (k_{d}/2)[(2J)^{2}-Q]dA+\oint [(k_{c}/2)
(2H+c_{0})^{2}+\mu ]dA+p\int dV,
\end{equation}
where $\mu$, $A$ and $V$ are surface tension, surface area and
volume enclosed by the vesicle, respectively. In this expression,
the term related to the Gaussian curvature disappears because its
integration $\oint KdA$ is an unimportant constant so that it is
omitted. It is easy to see that Eq.(\ref{freeeng}) is degenerated
to the free energy of closed lipid bilayer for $k_d=0$, and to the
free energy of solid shell with $\nu=1/2$ if $c_0=0$,
$k_d=C/(1-\nu^2)$ and $k_c=D$.

\section{\label{shsteqsec}The shape and in-plane strain equations of closed polymer vesicles}
In this section, we will give the shape and in-plane strain
equations of closed polymer vesicles from the first order
variation of free energy (\ref{freeeng}). The method has been
fully developed in Ref.\cite{tzcjpa}, and the key elements and
notations are shown in the Appendix.

If a point $\textbf{r}_0$ in a surface undergoing a displacement
$\textbf{u}$ to arrive at point $\textbf{r}$, we have $d\mathbf{u}
=d\mathbf{r}-d\mathbf{r}_{0}$ and naturally $\delta
_{i}d\mathbf{u} =\delta _{i}d\mathbf{r}$ ($i=1,2,3$).

If denote $d\mathbf{r}=\omega _{1}\mathbf{e}_{1}+\omega
_{2}\mathbf{e}_{2}$ and $d\mathbf{u} =\mathbf{U}_{1}\omega
_{1}+\mathbf{U}_{2}\omega _{2}$ with $|\mathbf{U}_{1}| \ll
1,|\mathbf{U}_{2}|\ll 1$, we can define the in-plane strains
\cite{wujk}:
\begin{eqnarray}
\varepsilon _{11} &=&\left[ \frac{d\mathbf{u}\cdot \mathbf{e}_{1}}{|d\mathbf{%
r}_{0}|}\right] _{\omega _{2}=0}\approx \mathbf{U}_{1}\cdot \mathbf{e}_{1},\label{epsl11} \\
\varepsilon _{22} &=&\left[ \frac{d\mathbf{u}\cdot \mathbf{e}_{2}}{|d\mathbf{%
r}_{0}|}\right] _{\omega _{1}=0}\approx \mathbf{U}_{2}\cdot \mathbf{e}_{2},\label{epsl22} \\
\varepsilon _{12} &=&\frac{1}{2}\left[ \left( \frac{d\mathbf{u}\cdot \mathbf{%
e}_{2}}{|d\mathbf{r}_{0}|}\right) _{\omega _{2}=0}+\left( \frac{d\mathbf{u}%
\cdot \mathbf{e}_{1}}{|d\mathbf{r}_{0}|}\right) _{\omega
_{1}=0}\right]
\approx \frac{1}{2}\left( \mathbf{U}_{1}\cdot \mathbf{e}_{2}+\mathbf{U}%
_{2}\cdot \mathbf{e}_{1}\right).\label{epsl12} \end{eqnarray}

Using $\delta _{i}d\mathbf{u} =\delta _{i}d\mathbf{r}$ and the
definitions of strains (\ref{epsl11})--(\ref{epsl12}), we can
obtain the leading terms of variational relations:
\begin{eqnarray} \delta _{i}\varepsilon _{11}\omega _{1}\wedge
\omega _{2} &=&\delta
_{i}\omega _{1}\wedge \omega _{2},\label{epsli11} \\
\delta _{i}\varepsilon _{12}\omega _{1}\wedge \omega _{2} &=&\frac{1}{%
2}[\omega _{1}\wedge \delta _{i}\omega _{1}+\delta _{i}\omega
_{2}\wedge
\omega _{2}],\label{epsli12} \\
\delta _{i}\varepsilon _{22}\omega _{1}\wedge \omega _{2}
&=&\omega _{1}\wedge \delta _{i}\omega _{2}.\label{epsli22}
\end{eqnarray}

From Eqs.(\ref{omvaratione11})--(\ref{detaomegaij}) and
(\ref{epsli11})--(\ref{epsli22}), we have:
\begin{eqnarray} \delta _{1}\mathcal{F}&=&\oint k_d[-d(2J)\wedge
\omega _{2}-\frac{\varepsilon _{11}d\omega _{2}-\varepsilon
_{12}d\omega _{1}}{2}+\frac{d(\varepsilon _{12}\omega
_{1}+\varepsilon _{22}\omega _{2})}{2}]\Omega
_{1},\label{deltamf1} \\
\delta _{2}\mathcal{F}&=& \oint k_d[d(2J)\wedge \omega
_{1}-\frac{\varepsilon _{12}d\omega _{2}-\varepsilon _{22}d\omega
_{1}}{2}-\frac{d(\varepsilon _{11}\omega _{1}+\varepsilon
_{12}\omega _{2})}{2}]\Omega
_{2},\label{deltamf2}\\
\delta
_{3}\mathcal{F}&=&\oint[k_c(2H+c_0)(2H^{2}-c_0H-2K)+k_c\nabla
^{2}(2H)\nonumber\\&&\qquad+ p-2H(\mu
+k_{d}J)-\frac{k_{d}}{2}(a\varepsilon _{11}+2b\varepsilon
_{12}+c\varepsilon _{22})]\Omega _{3}dA.\label{deltamf3}
\end{eqnarray} Thus the Euler-Lagrange equations corresponding to
the functional (\ref{freeeng}) are \begin{eqnarray}
&&k_d[-d(2J)\wedge \omega _{2}-\frac{1}{2}(\varepsilon
_{11}d\omega _{2}-\varepsilon _{12}d\omega
_{1})+\frac{1}{2}d(\varepsilon _{12}\omega
_{1}+\varepsilon _{22}\omega _{2})] =0,\label{shapecm1} \\
&&k_d[d(2J)\wedge \omega _{1}-\frac{1}{2}(\varepsilon _{12}d\omega
_{2}-\varepsilon _{22}d\omega _{1})-\frac{1}{2}d(\varepsilon
_{11}\omega
_{1}+\varepsilon _{12}\omega _{2})] =0, \label{shapecm2}\\
&& p-2H(\mu +k_{d}J)+k_c(2H+c_0)(2H^{2}-c_0H-2K)+k_c\nabla
^{2}(2H)\nonumber\\&&\qquad-\frac{k_{d}}{2}(a\varepsilon
_{11}+2b\varepsilon _{12}+c\varepsilon _{22}) =0.
\label{shapecm3}\end{eqnarray} Eqs.(\ref{shapecm1}) and
(\ref{shapecm2}) are called the in-plane strain equations because
they describe the in-plane strains of polymer vesicles under the
pressure $p$. Eq.(\ref{shapecm3}) is called the shape equation
because it describes the equilibrium shape of polymer vesicles
under the pressure $p$.

Obviously, if $k_d=0$, then Eqs. (\ref{shapecm1}) and
(\ref{shapecm2}) are two identities while Eq.(\ref{shapecm3}) is
degenerated into shape equation (\ref{shapelb}) of closed lipid
bilayers. Generally speaking, it is difficult to find the
analytical solutions to Eqs.(\ref{shapecm1})--(\ref{shapecm3}).
But it is easy to verify that
$\epsilon_{11}=\epsilon_{22}=\varepsilon$ (a constant),
$\epsilon_{12}=0$ satisfy Eqs.(\ref{shapecm1})--(\ref{shapecm3})
for a spherical vesicle with radius $R$ if the following equation
is valid:
\begin{equation}\label{spherecem1}
pR^{2}+(2\mu +3k_{d}\varepsilon)R+k_cc_0(c_0R-2)=0.
\end{equation}

\section{\label{stabsec}Mechanical stability of spherical polymer vesicles}
Now we will calculate the second order variation of functional
(\ref{freeeng}) and discuss the mechanical stability of a
spherical polymer vesicle.

In Ref.\cite{tzcjpa}, only the term $\delta _{3}^2\mathcal{F}$
related to the out-plane mode $\{\Omega_3\}$ is calculated. Here
we also consider the contribution of in-plane mode
$\{\Omega_1,\Omega_2\}$. Due to the notation of exterior
differential $d$ and Hodge star $\ast$, $\Omega_1$ and $\Omega_2$
can be expressed as
\begin{equation}\Omega _{1}\omega
_{1}+\Omega _{2}\omega _{2}=d\Omega +\ast d\chi\end{equation} by
two scalar potential functions $\Omega$ and $\chi$ for 2D manifold
\cite{westenholz}. Using
Eqs.(\ref{omvaratione11})--(\ref{detaomegaij}) and
(\ref{epsli11})--(\ref{epsli22}), we can calculate $\delta
_{1}^2\mathcal{F}$, $\delta _{2}^2\mathcal{F}$, $\delta
_{3}^2\mathcal{F}$, $\delta _{1}\delta _{2}\mathcal{F}$, $\delta
_{1}\delta _{3}\mathcal{F}$, and $\delta _{2}\delta
_{3}\mathcal{F}$ from Eqs.(\ref{deltamf1})--(\ref{deltamf3}) and
(\ref{spherecem1}) for spherical polymer membrane. Eventually, we
arrive at
\begin{equation}\delta ^2\mathcal{F}=\delta
_{1}^2\mathcal{F}+\delta _{2}^2\mathcal{F}+\delta
_{3}^2\mathcal{F}+2\delta _{1}\delta _{2}\mathcal{F}+2\delta
_{1}\delta _{3}\mathcal{F}+2\delta _{2}\delta
_{3}\mathcal{F}\equiv G_1+G_2,\end{equation} where
\begin{eqnarray} G_1&=&\oint\Omega
_{3}^{2}\{3k_{d}/R^{2}+(2k_cc_0/R^{3})+p/R\}dA\nonumber
\\&&+\oint\Omega
_{3}\nabla ^{2}\Omega _{3}\{k_cc_0/R+2k_c/R^{2}+ pR/2\}dA+\oint
k_c(\nabla
^{2}\Omega _{3})^{2}dA\nonumber \\
&&+\frac{3k_{d}}{R}\oint\Omega _{3}\nabla ^{2}\Omega
dA+k_{d}\oint\left( \nabla ^{2}\Omega \right) ^{2}dA+\frac{k_{d}}{2R^{2}}%
\oint\Omega \nabla ^{2}\Omega dA,\\
G_2&=&\frac{k_{d}}{4}\oint(\nabla ^{2}\chi )^{2}dA+\frac{k_{d}}{2R^{2}}%
\oint\chi \nabla ^{2}\chi dA. \end{eqnarray} If we take
$\kappa={k}_c/2$, $K=3k_d/2$, $\mu=k_d/2$, $w=\Omega_3$ and
$\Psi=\Omega$ in equations (6) and (7) of Zhang \textit{et al.}'s
paper \cite{zhangz}, then $G_1$ and $G_2$ correspond to
$F_1[w,\Psi]$ and $F_2[\chi]$ in that paper under the conditions
of $p=0$ and ${c}_0R=2$. Obviously, there is no coupling between
modes $\{\chi\}$ and $\{\Omega,\Omega_3\}$; but there is coupling
between in-plane mode $\{\Omega\}$ and out-of-plane mode
$\{\Omega_3\}$. We will show that in-plane modes have quantitive
effect on the stability of the cell membrane although they can not
qualitatively modify the results of Ref.\cite{tzcjpa}.

Because $G_2$ is obviously positive definite, we merely need to
discuss $G_1$. $\Omega_{3}$ and $\Omega$ in the expression of
$G_1$ can be expanded by spherical harmonic functions
\cite{wangzx} as $\Omega _{3} =\sum_{l=0}^{\infty
}\sum_{m=-l}^{m=l}a_{lm}Y_{lm}(\theta,\phi)$ and $\Omega
=\sum_{l=0}^{\infty }\sum_{m=-l}^{m=l}b_{lm}Y_{lm}(\theta,\phi)$
with $a_{lm}^{*} =(-1)^{m}a_{l,-m}$ and $b_{lm}^{*}
=(-1)^{m}b_{l,-m}$. It follows that \begin{eqnarray} G_{1}
&=&\sum_{l=0}^{\infty
}\sum_{m=0}^{l}2|a_{lm}|^{2}%
\{3k_{d}+[l(l+1)-2][l(l+1)k_c/R^{2}-k_cc_0/R-pR/2]\} \nonumber\\
&&-\sum_{l=0}^{\infty
}\sum_{m=0}^{l}\frac{3k_{d}}{R}l(l+1)(a_{lm}^{\ast
}b_{lm}+a_{lm}b_{lm}^{\ast })\nonumber\\ &&+\sum_{l=0}^{\infty }\sum_{m=0}^{l}\frac{k_{d}}{%
R^{2}}\left[ 2l^{2}(l+1)^{2}-l(l+1)\right] |b_{lm}|^{2}.
\end{eqnarray}

We find that if $p<p_{l}=\frac{3k_{d}}{\left[ 2l(l+1)-1\right] R}+\frac{2k_c[l(l+1)-c_0R]%
}{R^{3}}\quad(l=2,3,\cdots)$, then $G_{1}$ is positive definite,
i.e., the vesicle is stable. We must take the minimum of $p_l$ to
obtain the critical pressure:
\begin{equation}p_{c}=\min \{p_{l}\}=\left\{
\begin{array}{c}
\frac{3k_{d}}{11R}+\frac{2k_c[6-c_0R]}{R^{3}}<\frac{k_c[23-2c_0R]}{%
R^{3}},\quad (3k_{d}R^{2}<121k_c) \\
\frac{2\sqrt{3k_{d}k_c}}{R^{2}}+\frac{k_c}{R^{3}}(1-2c_0R),\quad
(3k_{d}R^{2}>121k_c)%
\end{array}%
\right.. \label{criticalpcm}
\end{equation}

But if we do not consider the in-plane mode $\{\Omega\}$, we will
obtain the critical pressure
\cite{tzcjpa}:\begin{equation}p_{c}=\left\{
\begin{array}{c}
\frac{3k_{d}}{2R}+\frac{2k_{c}(6-{c}_{0}R)}{R^{3}}<\frac{2k_{c}(10-{c}_{0}R)}{%
R^{3}},\quad (3k_{d}R^{2}<16k_{c}) \\
\frac{4\sqrt{3k_{d}k_{c}}}{R^{2}}+\frac{2k_{c}}{R^{3}}(2-{c}_{0}R),\quad
(3k_{d}R^{2}>16k_{c})
\end{array}%
\right.. \label{criticalpcmno}\end{equation}

Comparing Eq.(\ref{criticalpcm}) with (\ref{criticalpcmno}), we
find that in-plane mode have quantitive effect on the stability of
the polymer vesicles although they can not qualitatively modify
the result without considering it.

Now we test the validity of Eq.(\ref{criticalpcm}) by considering
two special cases. The first case, $k_d=0$, corresponds to lipid
bilayer. In this case, Eq.(\ref{criticalpcm}) is exactly reduced
to Eq.(\ref{critplipd}), the critical pressure for spherical lipid
bilayer. The second case, $c_0=0$, $k_d=Yh/(1-\nu^2)$ and
$k_c=Yh^3/[12(1-\nu^2)]$ with $\nu=1/2$, corresponds to the solid
shell with Young's modulus $Y$, Poisson ratio $\nu$, shell
thickness $h$. Under the condition of $h\ll R$,
Eq.(\ref{criticalpcm}) gives $p_c=(4/3)Yh^2/R^2$ that is exactly
the result of Eq.(\ref{critpshell}) with $\nu=1/2$. Thus we are
sure of the validity of Eq.(\ref{criticalpcm}).

Now we turn to the polymer vesicle consisting of polyelectrolytes.
In the experiment by Gao \emph{et al.} \cite{cgao}, its thickness
is $h\sim 20 nm$ which is much smaller than its radius $R\sim 2
\mu m$. The segment length is taken as 3 times of carbon-carbon
bond length, i.e., $b_0\sim 4.2$\AA. The number of segments per
polymer are about $N\sim 200$ due to the molecule weight 70000.
Thus $R_0\sim \sqrt{N}b_0\sim 60$\AA, which is less less than $R$.
Considering $k_d=4MhT$ and $k_c=MhR_0^2T/12$, we arrive at
$3k_{d}R^{2}/(121k_{c})=(12R)^2/(11R_0)^2\gg 1$, i.e.,
$3k_{d}R^{2}\gg 121k_{c}$. Under this condition,
(\ref{criticalpcm}) is reduced to
\begin{equation}
p_c=\frac{2MTR_0h}{R^2}.
\end{equation}
Above equation can explain the experimental result $p_c\sim
R^{-2}$ obtained by Gao \emph{et al.} But their result $p_c\sim
h^2$ is inconsistent with our theoretical result $p_c\sim h$. The
main reason for this discrepancy is that the membrane consisting
of polyelectrolytes is chemically instable if its layer number is
more than 10, which will disturb to test the exact relation
between critical pressure $p_c$ and thickness $h$.

\section{\label{Conclusion} Conclusion}
In above discussion, we briefly introduce the polymer statistics
and the correspondence principle (shown in Table \ref{correspond})
between it and path integral method in quantum mechanics. We
derive the entropy of a polymer confined in a curved surface and
the elastic energy of a membrane consisting of polymers by scaling
analysis. It is found that the elastic energy of the polymer
membrane has the form of the in-plane strain energy plus
Helfrich's curvature energy as shown in Eq.(\ref{elastic}).  The
elastic constants $k_d$, $k_c$, $\mu$ in the free energy are
obtained by discussing two simplified models: one is the polymer
membrane without in-plane strains and asymmetry between its two
sides, which is the counterpart of quantum mechanics in curved
surface; another is the planar rubber membrane with homogeneous
in-plane strains. The equations to describe equilibrium shape and
in-plane strains of polymer vesicles by osmotic pressure are
derived by taking the first order variation of the total free
energy (\ref{freeeng}) containing the elastic free energy, the
surface tension energy and the term induced by osmotic pressure.
The critical pressure (\ref{criticalpcm}), above which the
spherical polymer vesicle will lose its stability, is obtained by
taking the second order variation of the total free energy
(\ref{freeeng}). It is found that the in-plane mode $\{\Omega\}$
also plays important role in the critical pressure because it
couples with the out-of-plane mode $\{\Omega_3\}$.

We estimate that $p_c=\frac{2MTR_0h}{R^2}$ through the experiment
by Gao \textit{et al.} This result is qualitatively intermediate
between $p_{cl}=\frac{2k_{c}(6-c_{0}R)}{R^{3}}$ for lipid bilayer
and $p_{cs}=\frac{2Yh^2}{\sqrt{3(1-\nu^2)}R^2}$ for solid shell.
Therefore polymer vesicles possess the mechanical properties being
intermediate between Helfrich's fluid membranes and classical
solid shells. But is it reasonable to use the present theory to
polyelectrolyte membranes? We discuss two points: (i) The
derivation of the bending rigidity of the polymer membrane uses
results for the quantum mechanics of a particle constraint in a
curved surface. The results can only apply to polymers that are
much more constraint perpendicular to the membrane than their
lateral size. It seems that the polymer vesicles in the experiment
by Gao \textit{et al.} do not satisfy this condition since
$R_0\sim 6$ nm and $h\sim 20$ nm. But in fact, the vesicles
contain 10 layers and the thickness of each layer is about 2 nm
that is much small than $R_0$ and the total length of a polymer.
That is, the strong constraint in the normal direction of the
membrane is satisfied. (ii) The Gaussian chain model is used in
the present work. However, the polymers in polyelectrolyte
membranes may be nonideal. This is indeed a difficulty. But the
present theory is focused on the small deformations of polymer
vesicles, the model of ideal polymer should give approximate
results due to the lessons in classical theory of rubber
elasticity \cite{Treloar}. Additionally, we indeed obtain the
relation $p_c\sim R^{-2}$ observed by the experiment.

It is a nontrivial thing that we analysis the mechanical
stabilities directly from second order variations of free energy
(\ref{freeeng}). In the classical literature on stabilities of
shells, such as Ref.\cite{landau} and \cite{Pogorelov}, the
critical pressure (\ref{critpshell}) are obtained under a special
assumption of the instable mode that the concave part after
instability is the mirror image of its initial one. Therefore, the
present work implies that we can also analysis the mechanical
stabilities of solid shells directly from second order variations
of free energy (\ref{solidsh}) without the special assumption in
conventional literature.

It is well known that cell membranes contains lipid bilayers and
membrane skeleton. Our theory of polymer membranes may be
applicable for the membrane skeleton because it is also a
cross-linking structure at molecular levels. In the future, we can
turn to the elasticity of cell membranes after we fully studied
the elasticity of lipid bilayer and membrane skeleton.

\section*{Acknowledgement}
Z.C.T. acknowledges the help of Dr. R. An for her providing Ref.
\cite{Pogorelov} and useful communications with Prof. P. Pincus
\textit{et al.} in APCTP Focus Program on Biopolymers and
Membranes (POSTECH, Korea 2004).

\appendix

\section{Surface variation theory} Here we briefly retrospect the surface variation theory
originated in Ref.\cite{tzcpre} and fully developed in
Ref.\cite{tzcjpa}.

\begin{figure}[!htp]
\begin{center}
\includegraphics[width=6cm]{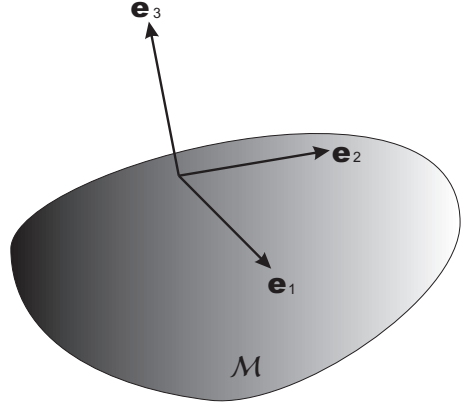}
\caption{\label{surfacem} Smooth and orientable surface $M$. we
can construct a right-hand orthonormal system
$\{\mathbf{e}_1,\mathbf{e}_2,\mathbf{e}_3\}$ at any point
$\mathbf{r}$ in the surface and call
$\{\mathbf{r};\mathbf{e}_1,\mathbf{e}_2,\mathbf{e}_3\}$ a moving
frame.}\end{center}
\end{figure}

We use a smooth and closed surface $\mathcal{M}$ in 3-dimensional
Euclid space $\mathbb{E}^3$ to represent a membrane. As shown in
Fig.~\ref{surfacem}, we can construct a right-hand orthonormal
system $\{\mathbf{e}_1,\mathbf{e}_2,\mathbf{e}_3\}$ at any point
$\mathbf{r}$ in the surface and call
$\{\mathbf{r};\mathbf{e}_1,\mathbf{e}_2,\mathbf{e}_3\}$ a moving
frame. The differential of the frame is denoted by
\begin{equation}\label{infiniter}\left\{
\begin{array}{l}
d\mathbf{r}=\omega_1\mathbf{e}_1+\omega_2\mathbf{e}_2,\\
\label{dei}d\mathbf{e}_i=\omega_{ij}\mathbf{e}_j\quad (i=1,2,3),
\end{array}\right.\end{equation} where $\omega_1$, $\omega_2$ and $\omega_{ij}=-\omega_{ji}$
$(i,j=1,2,3)$ are 1-forms. The structure equations of the surface
are \begin{eqnarray} \label{domega1}
d\omega_1&=&\omega_{12}\wedge\omega_2;\\
\label{domega2}
d\omega_2&=&\omega_{21}\wedge\omega_1;\\
\label{omega13} \omega_{13}&=&a\omega_{1}+b\omega_{2},\quad \omega_{23}=b\omega_{1}+c\omega_{2};\\
\label{domgaij} d\omega_{ij}&=&\omega_{ik}\wedge\omega_{kj}\quad
(i,j=1,2,3). \end{eqnarray} Readers should notice that the
operator ``$d$" is an exterior differential operator \cite{tzcjpa}
in this paper. The area element, mean curvature and Gaussian
curvature are respectively expressed as:
\begin{eqnarray}
&&dA=\omega_1\wedge\omega_2, \\
&&H=(a+c)/2 \label{meanh},\\
&&K=ac-b^2\label{gassianK}. \end{eqnarray}

If $\mathcal{M}$ undergoes an infinitesimal deformation such that
every point $\mathbf{r}$ in $\mathcal{M}$ has a displacement
$\delta\mathbf{r}$, we obtain a new surface
$\mathcal{M}'=\{\mathbf{r}'|\mathbf{r}'=\mathbf{r}+\delta\mathbf{r}\}$.
$\delta\mathbf{r}$ is called the variation of surface
$\mathcal{M}$ and can be expressed as \begin{eqnarray} \delta
\mathbf{r}=\delta_1\mathbf{r}+\delta_2\mathbf{r}+\delta_3\mathbf{r},\label{deltar}\\
\delta_i\mathbf{r}=\Omega_{i}\mathbf{e}_{i}\quad
(i=1,2,3),\label{deltari} \end{eqnarray} where the repeated
subindexes do not represent Einstein summation. Due to the
deformation of $\mathcal{M}$,
${\mathbf{e}_1,\mathbf{e}_2,\mathbf{e}_3}$ also change. We denote
the change as \begin{equation} \delta_l \mathbf{e}_{i}=\Omega
_{lij}\mathbf{e}_{j},\quad \Omega _{lij}=-\Omega
_{lji}.\end{equation}

Using the commutativity between $\delta_i$ $(i=1,2,3)$ and $d$, we
obtain the fundamentally variational identities of the move frame
\cite{tzcjpa}: \begin{eqnarray}
&&\delta_1 \omega _{1} =d\Omega _{1}-\omega _{2}\Omega _{121},\label{omvaratione11}\\
&&\delta_1 \omega _{2} =\Omega _{1}\omega _{12}-\omega _{1}\Omega _{112},\label{omvaratione12}\\
&&\Omega _{113}=a\Omega _{1},\quad\Omega _{123}=b\Omega
_{1};\label{omvaratione13}\\
&&\delta _{2}\omega _{1} =\Omega _{2}\omega _{21}-\omega _{2}\Omega _{221},\label{omvaratione21} \\
&&\delta _{2}\omega _{2} =d\Omega _{2}-\omega _{1}\Omega _{212},\label{omvaratione22} \\
&&\Omega _{213} =b\Omega _{2},\quad\Omega _{223}=c\Omega
_{2};\label{omvaratione23}\\
&&\delta_3 \omega _{1} =\Omega _{3}\omega _{31}-\omega _{2}\Omega
_{321},
\label{detaomega1} \\
&&\delta_3 \omega _{2} =\Omega _{3}\omega _{32}-\omega _{1}\Omega
_{312},
\label{detaomega2} \\
&&d\Omega _{3} =\Omega _{313}\omega _{1}+\Omega _{323}\omega _{2};
\label{domega3}\\
&&\delta_l \omega _{ij}=d\Omega _{lij}+\Omega _{lik}\omega
_{kj}-\omega _{ik}\Omega _{lkj}.\label{detaomegaij}\end{eqnarray}

\newpage

\begin{table*}[!htp]\caption {\label{correspond}The correspondence
principle between polymer statistics and path integral method in
quantum mechanics.}
\begin{ruledtabular}
\begin{tabular}{rl}
Quantum Mechanics & Polymer Statistics\\ \hline
time $t$ & the number of segments $N$\\
$i/\hbar$ & $-\beta=-1/T$\\
Lagrangian
$\hat{L}=(m/2)(d\mathbf{r}/d\tau)^2-V[\mathbf{r(\tau)}]$& energy
$E=(\eta/2) (d\mathbf{R}_n/dn)^2+U(\mathbf{R}_n)$\\
mass $m$ &  $\eta=3/(\beta b_0^2)$\\
potential $V(\mathbf{r(\tau)})$ & $-U(\mathbf{R}_n)$\\
 propagator
$\hat{K}=\int\exp[(i/\hbar)\int_0^t\hat{L}d\tau]D[\mathbf{r}(\tau)]$
& partition function $Z=\int\exp[-\beta\int_0^NEdn]D[\mathbf{R}_n]$\\
Hamiltonian $\hat{H}=-\frac{\hbar^2}{2m}\nabla^2+V$ &
$\hat{P}=\frac{1}{2\eta\beta^2}\nabla^2-U$\\
$i\hbar\partial \hat{K}/\partial t=\hat{H}\hat{K}, (t>0)$ &
$(1/\beta)\partial Z/\partial N=\hat{P}Z$
\end{tabular}
\end{ruledtabular}
\end{table*}
\end{document}